\documentclass[aps,prb,twocolumn,superscriptaddress,showpacs]{revtex4}
\pdfoutput=1

\usepackage{graphicx}
\usepackage{dcolumn}
\usepackage{bm}
\usepackage{latexsym}
\usepackage{amssymb}
\usepackage{amsfonts}
\usepackage{amsmath}

\begin{document}
\title{Andreev reflection in graphene nanoribbons}
\date{\today}

\author{Diego Rainis}
\affiliation{NEST-CNR-INFM and Scuola Normale Superiore, I-56126 Pisa, Italy}
\author{Fabio Taddei}
\affiliation{NEST-CNR-INFM and Scuola Normale Superiore, I-56126 Pisa, Italy}
\author{Fabrizio Dolcini}
\affiliation{NEST-CNR-INFM and Scuola Normale Superiore, I-56126 Pisa, Italy}
\author{Marco Polini}
\affiliation{NEST-CNR-INFM and Scuola Normale Superiore, I-56126 Pisa, Italy}
\author{Rosario Fazio}
\affiliation{International School for Advanced Studies (SISSA), 
             via Beirut 2-4, I-34014 Trieste, Italy}
\affiliation{NEST-CNR-INFM and Scuola Normale Superiore, I-56126 Pisa, Italy}

\begin{abstract}
We study Andreev reflection in graphene nanoribbon/superconductor hybrid junctions. 
By using a tight-binding approach and the scattering formalism
we show  that finite-size effects lead to notable differences    
with respect to the bulk graphene case. At subgap voltages, conservation of pseudoparity, a quantum number characterizing the ribbon states, yields either a suppression of Andreev reflection  when the ribbon has an even number of sites in the transverse direction or 
perfect Andreev reflection when the ribbon has an odd number of sites. In the former case  the suppression of Andreev reflection 
induces an insulating behavior even when the junction is biased; electron conduction can however 
be restored by applying a gate voltage. 
\end{abstract}


\maketitle

\section{Introduction}

Graphene, a flat monolayer of Carbon atoms arranged in a two-dimensional (2D)  honeycomb lattice, is a newly realized 2D electron system~\cite{reviews} which has attracted a great deal of interest because of the novel physics which it exhibits and because of its potential as a new material for electronic technology. One of the fascinating aspects of this system is that in most cases its potentialities for device applications are intimately related with fundamental aspects of quantum mechanics.  
Many of the transport properties of graphene which are at the heart of the design of new functional nanostructures originate from conservation laws of certain quantum numbers. For example, the fact that electrostatic barriers in graphene are perfectly transparent to electrons scattering with angles close to normal incidence (Klein-paradox) is  explained~\cite{klein} in terms of conservation of pseudospin (the sublattice degree-of-freedom necessary to describe graphene's non-Bravais honeycomb lattice~\cite{reviews}).

Early investigations on transport properties of graphene have analyzed current and noise in the presence of normal leads. 
Recent studies~\cite{beenakker_prl_2006,Ludwig,Linder,Levy-Yeyati,Cayssol,Prada,gen_ref} have pointed out that  novel interesting phenomena arise when graphene is interfaced to a superconductor~(SC). 
In the seminal papers by Beenakker and coworkers~\cite{beenakker_prl_2006} it has been shown that the peculiar band structure of graphene  gives rise to the appearance of  specular  Andreev reflection~(AR)~\cite{andreev_JEPT_1964}, a novel type of  AR that is absent in ordinary metal/SC interfaces. These studies paved the way to experimental investigations of the proximity effect~\cite{Bouchiat,Miao} 
and of supercurrent flow~\cite{Heersche,Du} in graphene. On the theoretical side the results of Ref.~\onlinecite{beenakker_prl_2006} have been extended in a number of ways, {\em e.g.} to graphene bilayers~\cite{Ludwig} and to the case of interfaces with $d$-wave SCs~\cite{Linder}, whereas further studies have focused on the subgap structure of SC-graphene-SC junctions~\cite{Levy-Yeyati}, on 
crossed AR~\cite{Cayssol}, and on magneto-transport~\cite{Prada}.

Most of the theoretical analysis carried out so far describe the graphene sheet as an infinite (or semi-infinite) 2D plane, and identify two energy scales relevant for transport, namely the superconducting gap $\Delta_0$ and the potential difference between the normal side and the superconducting side, which allows, for instance, to switch from the regime of ordinary AR to the one of specular AR.
In graphene ribbons~\cite{GNR_exp}, however, the finite size of the sample yields an additional energy scale $\delta$, characterizing the mean energy spacing between the ribbon bands. Since the typical ribbon size varies from 
$10~{\rm nm}$ up to $1~\mu{\rm m}$~\cite{GNR_exp}, $\delta$ can range from $300~{\rm  meV}$ down to $3~{\rm  meV}$, and it is thus 
larger than (or of the same order of) the typical superconducting gap $\Delta_0 \lesssim 1~{\rm meV}$. As a consequence, $\delta$ is expected to play an important role in electron transport and AR in graphene.

\begin{figure}[t!]
\begin{center}
\includegraphics[angle=-90, width=0.9\linewidth]{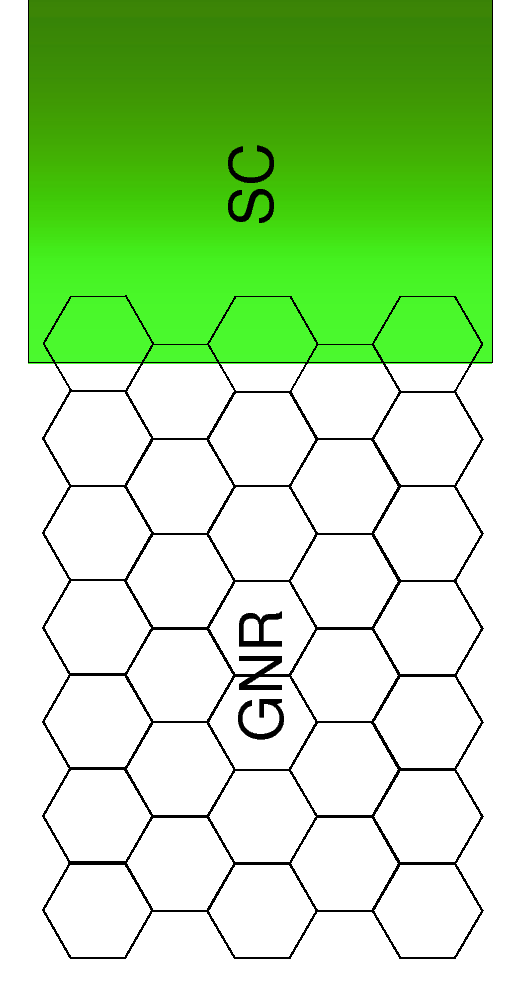}
\end{center}
\caption{(Color online) A hybrid junction between a zigzag graphene nanoribbon and a superconductor.\label{fig:one}}
\end{figure}

In the present work we address this problem by analyzing electronic transport through a hybrid junction between 
a Graphene NanoRibbon~(GNR)~\cite{GNR_theory,Brey-Fertig,onipko} and a SC, as sketched in Fig.~\ref{fig:one}. 
We consider the case of a GNR with zigzag edges, 
which has been shown to represent fairly well the behavior of an arbitrarily shaped edge~\cite{akhmerov_beenakker}. 

Novel effects emerge due to the finite size of the GNR. We find that AR is strongly affected by the conservation of {\it pseudoparity},  
a quantum number characterizing the GNR eigenstates, which depends on whether the number $N_W$ of  sites along the transverse direction is even or odd. 
In particular, while in GNRs with   odd $N_W$ the AR coefficient is unity for subgap voltages, in GNRs with an even $N_W$ AR is totally suppressed. In the latter case AR can be restored to a finite value by applying a gate voltage. 
Even-odd effects with the same physical origin have been also found in normal transport through graphene p-n junctions (valley-valve effect)~\cite{wakabayashi_ijmpB_2002,akhmerov_prb_2008,grosso_prb_2008}.

The contents of the paper are briefly described as follows. In Sect.~\ref{sect:model} we introduce the tight-binding 
Hamiltonian that we use to model the GNR  and  provide 
analytical expressions  for the GNR eigenfunctions at arbitrary energy, briefly discussing how transport properties of the hybrid GNR/SC junction are calculated. In Sect.~\ref{sect:results} we report and discuss our results. 
Finally, in Sect.~\ref{sect:conclusions} we summarize our main conclusions.

\section{Model Hamiltonian, GNR eigenfunctions, and transport properties}
\label{sect:model}

We model the hybrid GNR/SC system by means of a tight-binding approach.
In order to interpret the results presented below, we start by describing the eigenstates in the ``normal" side of the junction (the GNR side). 
The GNR Hamiltonian ${\cal H}$ is a matrix with elements
${\cal H}_{{\bm i} {\bm j}}=-\gamma \delta_{\langle {\bm i}, {\bm j} \rangle}+U \delta_{{\bm i} {\bm j}}$, where
 $\gamma \simeq 2.8$~eV is the hopping energy between nearest-neighbor sites $\langle {\bm i}, {\bm j} \rangle$ 
on the honeycomb lattice, and $U$ accounts for a constant electrostatic energy, which can be controlled by a gate voltage. 

\begin{figure}[t!]
\begin{center}
\includegraphics[width=0.8\linewidth]{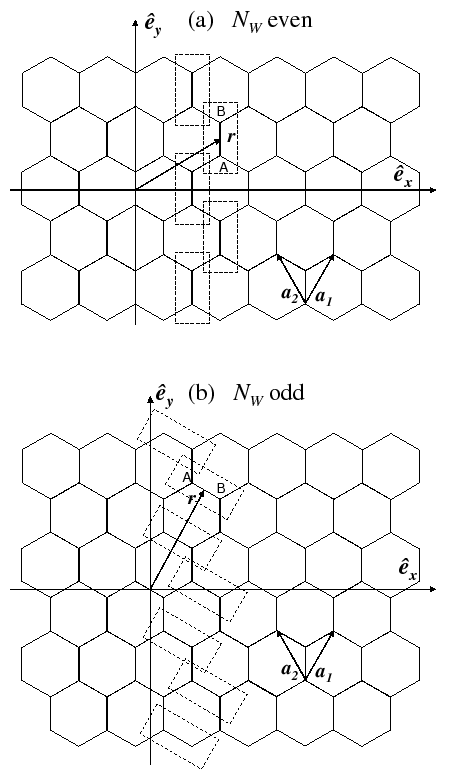}
\end{center}
\caption{Coordinates and elementary cells in GNRs with (a) an even number of sites in the transverse direction 
$\hat{\bm e}_{\bm y}$ 
and (b) an odd number of sites in the $\hat{\bm e}_{\bm y}$ direction.\label{fig:two}}
\end{figure}

Denoting by ${\bm a}_1$ and ${\bm a}_2$ the honeycomb-lattice basis, each unit cell is labeled by a vector 
${\bm r}= n_1 {\bm a}_1+n_2 {\bm a}_2$, with $n_1$ and $n_2$ integers. 
The origin is located at the geometrical center of the GNR and $N_W$ is the number of sites in the transverse direction~$\hat{\bm e}_{\bm y}$. A suitable shape of the cells turns out to depend on whether $N_W$ is even or odd, 
as depicted in Fig.~\ref{fig:two}. It can be shown that the (unnormalized)   eigenfunctions of a zigzag GNR at a cell 
${\bm r}=(x,y)$ are given by~\cite{footnote_solution}
\begin{equation}
\Psi_{A(B),s}(x,y)= e^{i k_x x_{A(B)}(x)} \Phi_{A(B),s}(y) \label{eigen-prel}
\end{equation}
where $x_{A(B)}(x)$ is the coordinate of the $A(B)$ site of the cell along the longitudinal direction~$\hat{\bm e}_{\bm x}$, $\Phi_{A(B),s}$ is the wavefunction along the transverse direction at $A(B)$, and $s=\uparrow,\downarrow$ is a spin label. Explicitly, $x_{A(B)}=x$ if $N_W$ is even, whereas 
$x_{A(B)}=x \mp a \sqrt{3}/4$ if $N_W$ is odd ($a \simeq 1.42$~\AA~is the Carbon-Carbon distance). Furthermore
\begin{equation}\label{eigen}
\left( 
\begin{array}{c} \Phi_{A,s}(y) \\ \Phi_{B,s}(y) 
\end{array} \right) = 
\left( 
\begin{array}{c} 
 \sin[k_y~(W/2+ y) ] \\ \eta~\sin[k_y~(W/2- y) ]
\end{array} \right)
\end{equation}
where $W\doteq 3a~{\rm int}[(N_W+1)/2]$ approximately coincides with the width of the GNR, 
with ${\rm int}[x]$ denoting the integer part of $x$.
The eigenvalues related to (\ref{eigen-prel}) read
\begin{equation}
\varepsilon = U + \eta~\gamma~\Re e~f(k_x a \sqrt{3},3 k_y a) \label{eigval}
\end{equation}
where 
\begin{equation}
f(\xi,\kappa)=e^{i \kappa N_W/2}[1+2e^{i\kappa/2}\cos{(\xi /2)}]~,
\end{equation}
and $k_y$ is the solution of $\Im m~f(k_x a \sqrt{3},3 k_y a)=0$ (see also Ref.~\onlinecite{onipko}). 
The quantity $k_x$ ($k_y$) is the component of the momentum along   
$\hat{\bm e}_{\bm x}$~($\hat{\bm e}_{\bm y}$). 
Importantly, beside the momentum, the eigenstates defined in Eqs.~(\ref{eigen-prel})-(\ref{eigen}) exhibit an additional quantum number 
$\eta=\pm 1$, which determines whether the transverse cell  wavefunction $(\Phi_A+\Phi_B)/2$ is even or odd with respect to the 
longitudinal axis of the GNR. One can therefore term $\eta$ as {\it pseudoparity}. 
Note that, in contrast to what happens in bulk graphene, the states~(\ref{eigen-prel})-(\ref{eigen}) are not eigenstates of the pseudospin-projection operator 
${\bm \sigma}\cdot {\hat {\bm n}}$ for any unit vector ${\hat {\bm n}}$.

\begin{figure}[t!]
\begin{center}
\includegraphics[width=0.85\linewidth]{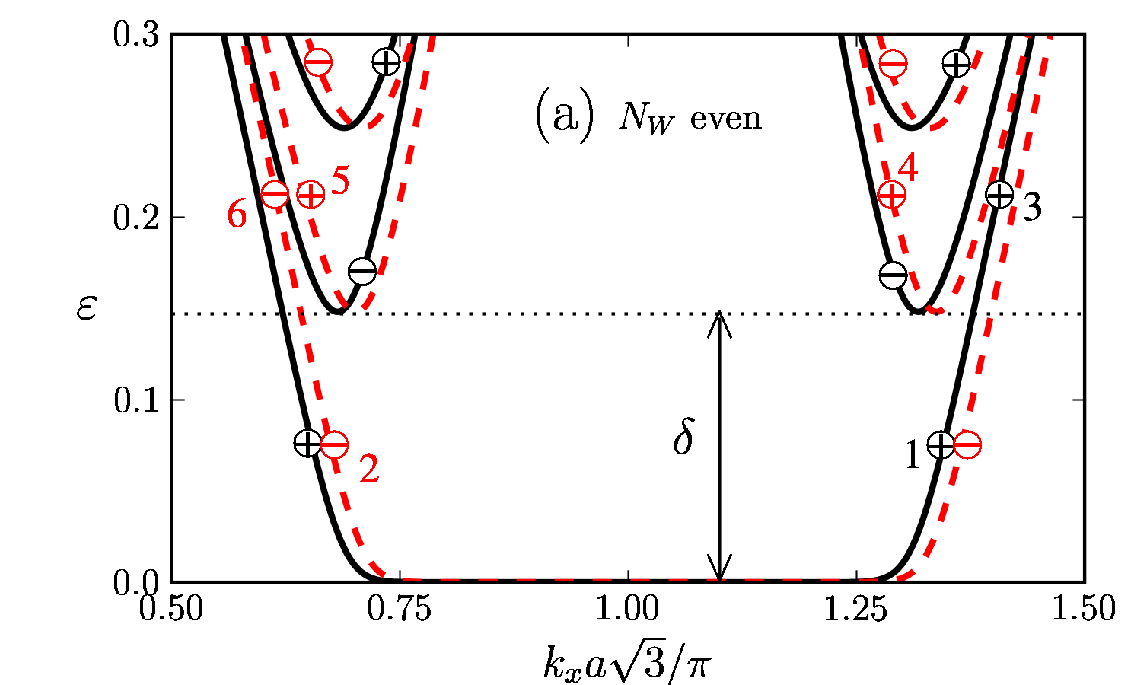}
\includegraphics[width=0.85\linewidth]{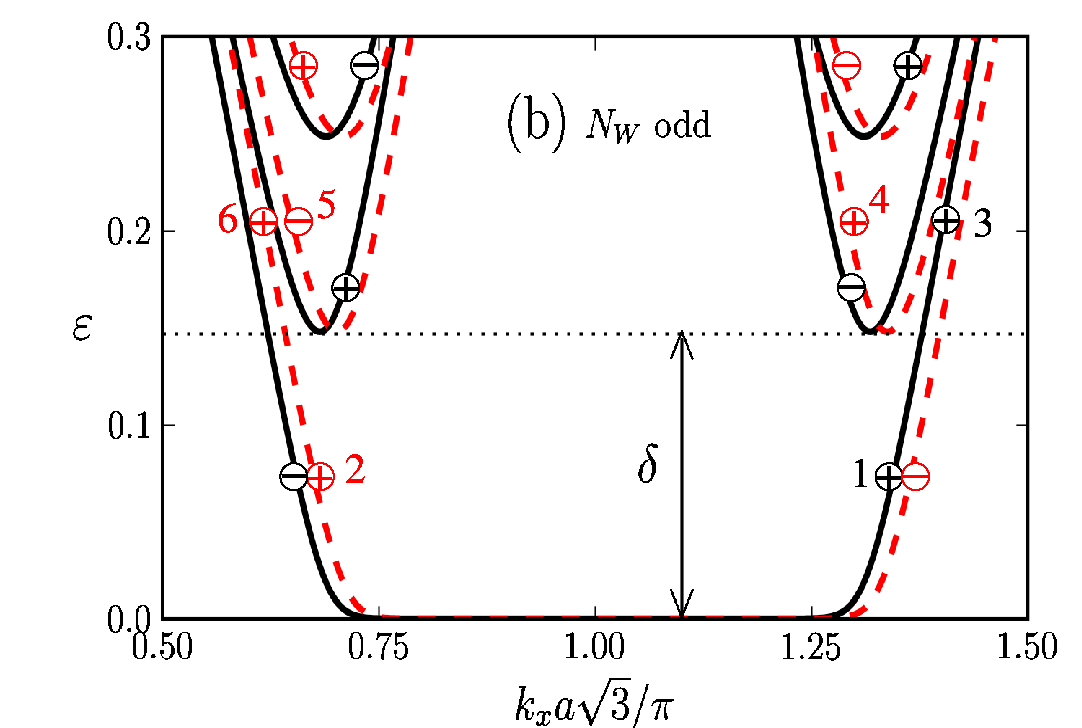}
\caption{(Color online) The band structure of a GNR 
as a function of the longitudinal wavevector $k_x$. Only the three lowest   bands are illustrated. 
Energies are in units of $\gamma$. The solid (dashed) lines represent the particle (hole) bands. In this figure  
$U=0$ (zero gate voltage). In this case particle and hole bands are degenerate (for the sake of clarity the hole bands have been slightly shifted to the right). The signs $+$ and $-$ inside the circles indicate the value $\eta$ of pseudoparity for each band. (a) $N_W$ is even. 
In the range $0< \varepsilon < \delta$ no AR is possible, since left-moving hole states ``2" have pseudoparity opposite to the one of incoming right-moving particle states ``1". In contrast, for $\varepsilon> \delta$, left-moving hole states ``4" and ``5" with the same pseudoparity as incoming particle states ``3" are available, and AR is finite. (b) $N_W$ is odd. In this case AR is possible in the range $0< \varepsilon < \delta$ since left-moving hole states ``2" have the same pseudoparity of incoming right-moving particle states ``1". 
For $\varepsilon> \delta$ left-moving hole states ``4" and ``6" with the same pseudoparity as incoming particle states ``3" are available 
and AR is again possible. Note that left-moving hole state ``5" in this case has pseudoparity opposite to the one of incoming right-moving particle states ``3".\label{fig:three}}
\end{center}
\end{figure}

The eigenvalues $\varepsilon$ as functions of~$k_x$ describe the band structure of the zigzag GNR, 
where two bands exhibit dispersionless zero modes and the other ones are somewhat reminiscent of the bulk-graphene Dirac cones~\cite{reviews}, as illustrated in Fig.~\ref{fig:three} for the case $U=0$. Energies are measured with respect to the Fermi level, and solid (dashed) lines refer to particles (holes) bands, which are degenerate for this particular case. The value of the pseudoparity $\eta$, 
shown inside the circles, alternates from a particle band to the next one, and takes opposite values in particle and hole bands. Importantly, in the range $k_x a \sqrt{3} \in [0, \pi]$, the pseudoparity of a GNR with even $N_W$ is opposite to the one for the case with odd $N_W$.

\begin{figure}[t!]
\begin{center}
\includegraphics[width=0.85\linewidth]{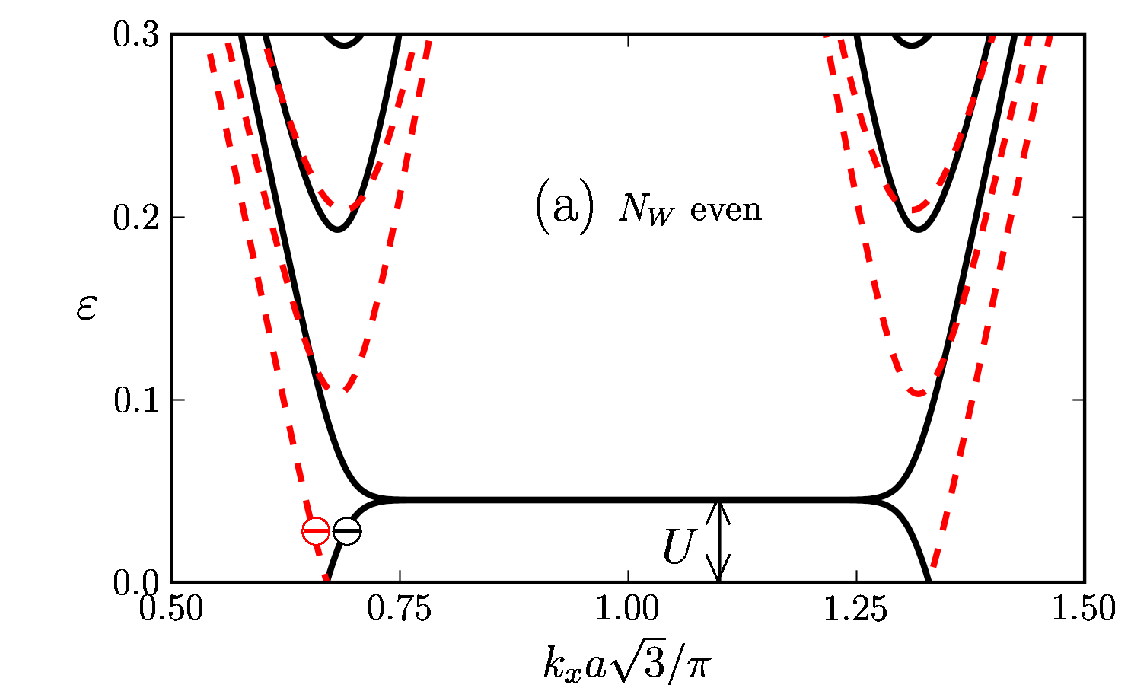}
\includegraphics[width=0.85\linewidth]{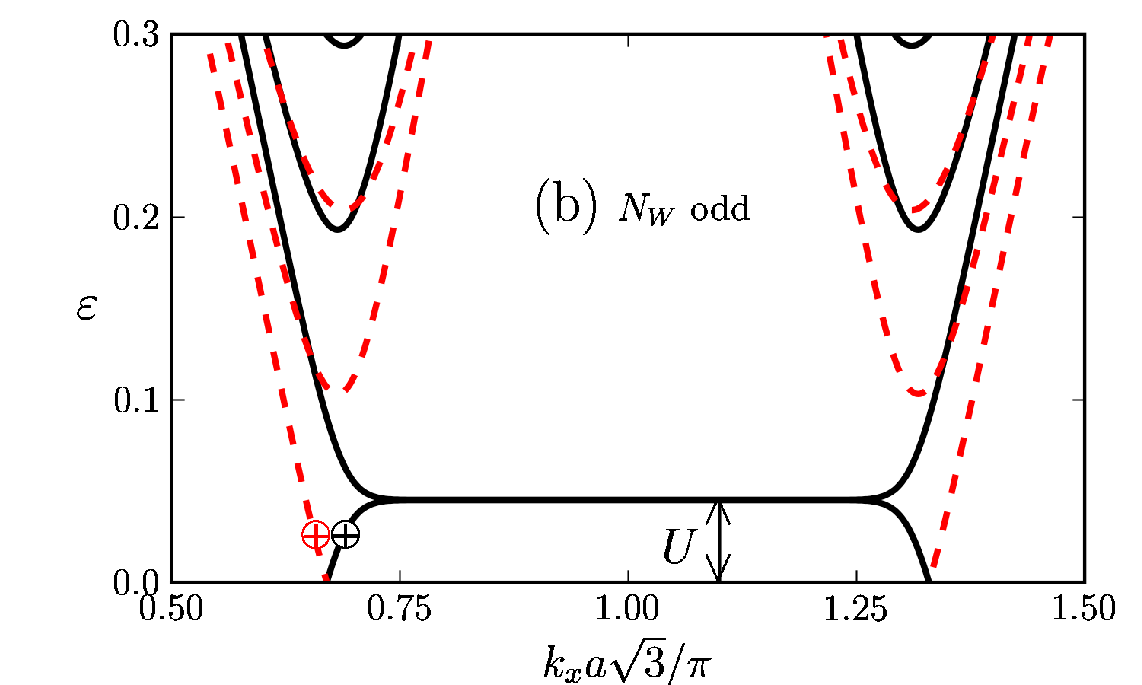}
\caption{(Color online) Same as in Fig.~\ref{fig:two} but for $U\neq 0$ (finite gate voltage). 
In this case for $\varepsilon <U$ incoming particle states and outgoing hole states have the same pseudoparity, and AR is allowed.\label{fig:four}}
\end{center}
\end{figure}

The energy separation $\delta$ between the first band and the zero-mode energy (Dirac level) is given by 
$\delta \simeq 9 \pi \gamma/(6N_W-4)$ for even $N_W\gg 1$ or 
$\delta \simeq 3\pi \gamma/(2N_W-2)$ for odd $N_W\gg 1$~\cite{Beenakker_bands}. 
A finite $U$ shifts the energy of the Dirac level away from the Fermi level, as shown in Fig.~\ref{fig:four}, breaking particle-hole degeneracy. 

The presence of the superconducting electrode is accounted for by the Bogoliubov-de Gennes Hamiltonian~\cite{degennes}
\begin{equation}\label{HBdG}
{\cal H}_{\rm BdG} = \left ( \begin{array}{cc} {\cal H}- \varepsilon_{\rm F} & \Delta \\ \Delta^* & \varepsilon_{\rm F} -{\cal H}^* \end{array} \right )~,
\end{equation}
where ${\cal H}$ is the particle Hamiltonian,   and 
$\varepsilon_{\rm F}$ is the Fermi level of the GNR/SC junction at equilibrium. 
The SC ($s$-wave) order parameter, which couples  particles and holes, is described by a matrix $\Delta$ with entries 
$\Delta_{{\bm i} {\bm j}}=\Delta_{\bm i} \delta_{{\bm i} {\bm j}}$. Here $\Delta_{\bm i}$ is taken to vary smoothly across the junction, 
between its maximum value  $\Delta_0$ in the bulk of the SC and zero in the bulk of the GNR~\cite{footnote_steplike}. 

Within the scattering approach~\cite{Datta} the zero-temperature differential conductance is given by~\cite{smatrix} 
\begin{equation}\label{G}
G(V)=\frac{4e^2}{h}
\left[N(V)-R(V)+R_{\text{A}}(V) \right]~,
\end{equation} 
where $V$ is the bias voltage applied across the junction.
In Eq.~(\ref{G}) the prefactor $4$ is due to spin and valley degeneracies, whereas $N(V)$, $R(V)$ and $R_{\text{A}}(V)$ are the number of transverse propagating modes (open channels) available at energy $eV$ measured from the Fermi level $\varepsilon_{\rm F}$, the normal reflection coefficient, and the AR coefficient, respectively. Below the gap ($eV < \Delta_0$) only AR processes can contribute to the conductance, since  the unitarity of the scattering matrix implies that $N-R=R_{\rm A}$ in this range.

\section{Results and discussion}
\label{sect:results}

We present  our   numerical results  focusing on the physically realistic regime $\delta \gg \Delta_0$. We first discuss the case where the Fermi level lies at the Dirac level ($U=0$) and then the case with a finite gate voltage $U\neq 0$. 

\subsection{Zero gate voltage}

\begin{figure}[t!]
\begin{center}
\includegraphics[width=0.8\linewidth, clip=]{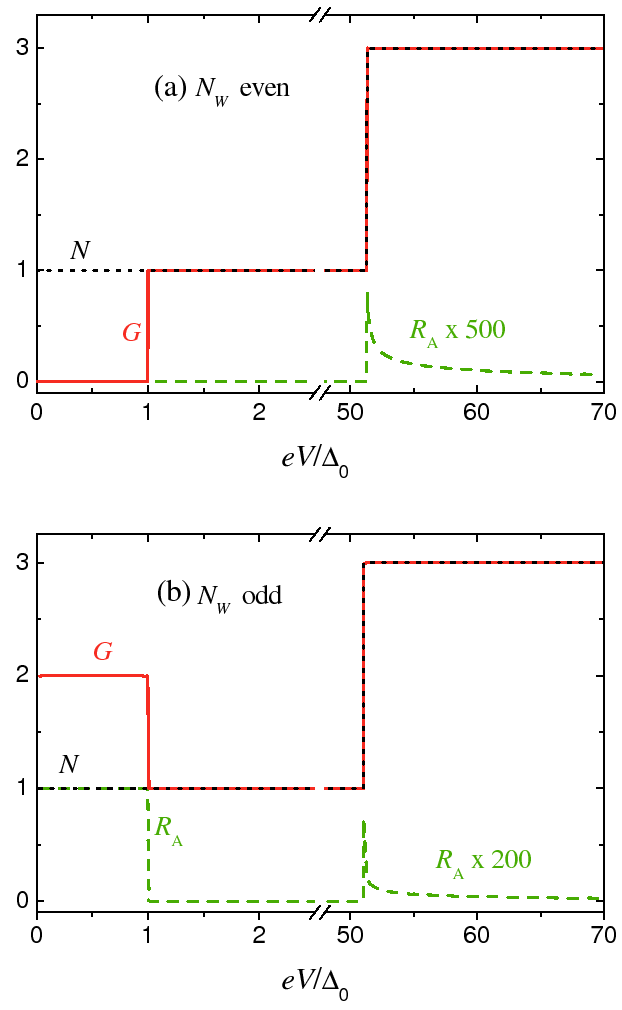}
\caption{(Color online) Transport properties of a GNR/SC interface. 
The differential conductance $G$ in units of  $4e^2/h$ (solid line), 
the total Andreev-reflection coefficient $R_{\rm A}$ (dashed line), and the number $N$ of open channels (dotted line) 
are shown as functions of the applied bias $eV$ (in units of $\Delta_0$). Note that the horizontal axis has been broken in two ranges. The GNR has a width of about $50~{\rm nm}$, which corresponds to $\delta \sim 52~{\rm meV}= 52~\Delta_0$ (the superconducting gap has been fixed at the value $\Delta_0 =1~{\rm meV}$). (a) $N_W=250$. The AR coefficient is non-zero only for $eV > \delta$, and has been multiplied by a factor $500$ for clarity. (b) $N_W=251$. The AR coefficient is unity for $eV < \Delta_0$ and drops abruptly to zero for  $eV > \Delta_0$, remaining zero up to $eV=\delta$.  The AR coefficient is again finite for $eV > \delta$, and has been multiplied by a factor $200$ for clarity.
\label{fig:five}}
\end{center}
\end{figure}

In Fig.~\ref{fig:five}(a) we plot $G$, $R_{\rm A}$, and $N$, as functions of the bias voltage applied across the junction, for a GNR with even $N_W$, and in Fig.~\ref{fig:five}(b) we plot the same quantities for a GNR with odd $N_W$. 
In the subgap regime $eV<\Delta_0$ a striking dependence   on the  number of sites arises. While for even $N_W$ the AR coefficient $R_{\rm A}$ {\it vanishes}, the opposite occurs for odd $N_W$, where $R_{\rm A}$ is unity. Because for $eV<\Delta_0$ electron transport is possible only by virtue of AR processes, 
the conductance $G$  also vanishes in this voltage range for even $N_W$. In contrast, for odd $N_W$, $G$ takes its maximum value in the same voltage range. In the regime $\Delta_0 < eV < \delta$, where quasi-particle transmission becomes possible, $R_A$  
drops abruptly to zero also for odd $N_W$. This is due to the fact that the transmission processes, which are characterized by a small momentum transfer (intra-valley), largely dominate over AR processes, which instead involve large momentum transfer (inter-valley). 
We thus find a finite conductance with a value   equal to the number $N$ of open channels (in units of $4e^2/h$). Finally, for $eV>\delta$, $R_A$ is finite but quite small in both cases (for clarity $R_A$ has been multiplied by an enhancement factor), since voltages in this range are well above the superconducting gap. 

These features ($R_{\rm A}=0$ for $eV<\delta$ if $N_W$ is even and $R_{\rm A}=1$ for $eV<\Delta_0$ if $N_W$ is odd) 
can be understood in terms of a pseudoparity superselection rule for the scattering states of the GNR. 
Let us consider a right-moving incoming electron (labeled by ``1'' in Fig.~\ref{fig:two}) with an energy lying between $\varepsilon_{\rm F}$ 
and the bottom of the second band. 
The only hole state available for an AR process is the one labeled by ``2'', which is characterized 
by a pseudoparity $\eta$ that is (i) {\it opposite} to the one of the incoming electron for even $N_W$ and (ii) {\it equal} 
to the one of the incoming electron for odd $N_W$. 

A superconducting parameter that is constant along the transverse direction 
cannot couple states with opposite pseudoparity ({\it i.e.} the pseudoparity of an electron impinging onto the SC interface cannot be flipped). 
Thus for even $N_W$ the AR process is forbidden and $R_A$ vanishes, whereas for odd $N_W$ this process is allowed 
and $R_A$ is finite. Notice that in the latter case normal reflection is forbidden by pseudoparity conservation, yielding $R_A=1$ in the subgap regime. For $\Delta_0 < eV  < \delta$ normal transmission is the dominating  process, so that $R_A$ is strongly suppressed for  both   even and odd $N_W$. A finite AR is restored for $eV>\delta$.   Indeed in this case intra-valley scattering into hole states with the appropriate pseudoparity
[``4''  in Figs.~\ref{fig:three}(a) and   Fig.~\ref{fig:three}(b)] is available. 
This superselection rule is the ribbon counterpart of the bulk Klein-paradox selection rules based on pseudospin~\cite{klein} and also explains  the valley-valve effect in p-n junctions in GNRs~\cite{wakabayashi_ijmpB_2002,akhmerov_prb_2008,grosso_prb_2008}.   

\subsection{Finite gate voltage}

\begin{figure}[t!]
\begin{center}
\includegraphics[width=0.8\linewidth, clip=]{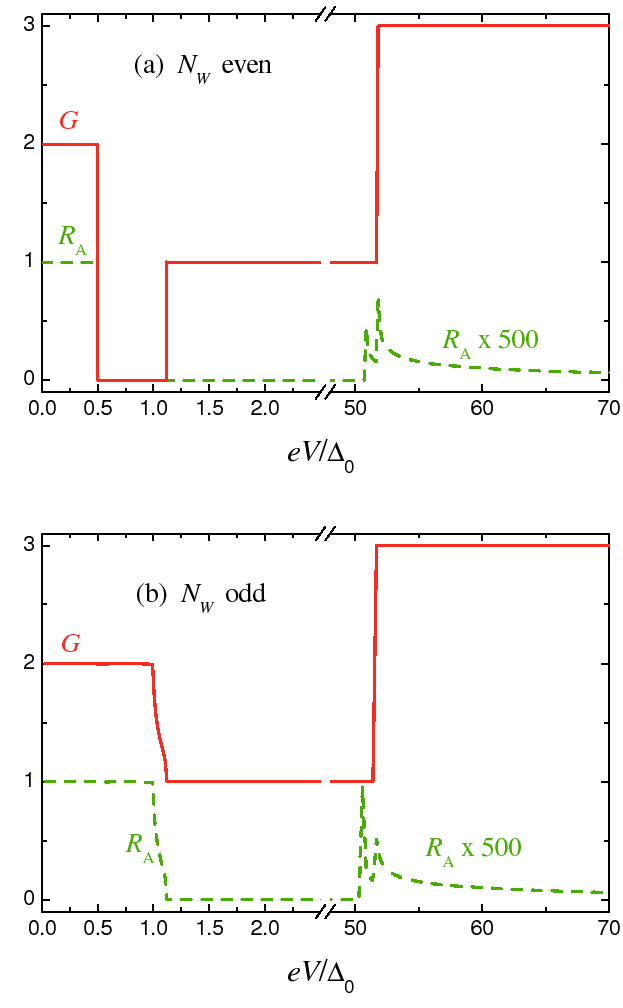}
\caption{(Color online) Same as in Fig.~\ref{fig:five} but for $U = 0.5~\Delta_0$ (in this figure we have not 
plotted the number of open channels). (a) $N_W$ even. In this case the AR coefficient
is restored to a finite value for $eV<U$, while it remains zero for $U< eV<\delta$. (b) $N_W$ odd. In this case 
no big qualitative changes are seen to occur by switching on a finite value of $U$.
\label{fig:six}}
\end{center}
\end{figure}

Let us now consider the case $U \neq 0$ (and $U \ll \delta$). In Fig.~\ref{fig:six} 
we plot our results for $G$, $R_{\rm A}$, and $N$. 

A comparison between Fig.~\ref{fig:five}(a) and Fig.~\ref{fig:six}(a) shows that the application of a gate voltage has dramatic consequences on electron conduction in GNRs with even $N_W$. This is evident from Fig.~\ref{fig:four}(a). AR processes are allowed by pseudoparity conservation in the bias range $eV < U$, while they are forbidden for $U< eV < \delta$. This fact is reflected in the  results shown in  Fig.~\ref{fig:six}(a), 
where the differential conductance is finite (and equal to $8e^2/h$) for $eV < U$ and vanishes in the range~\cite{footnote_sqrt} $U < eV  < \sqrt{ U^2 + \Delta^2_0}$. For subgap voltages electron conduction can be switched on and off by the application of a gate voltage. The junction between an even-$N_W$ GNR and a SC exhibits the operational behavior of an electron transistor based on the pseudoparity conservation law.

In  contrast, for odd $N_W$ the application of a gate voltage is expected to have 
no major qualitative impact on the transport properties of a junction 
between a GNR with odd $N_W$ and a SC. Once again, this expectation can be traced back 
to the pseudoparity quantum number, as depicted in Fig.~\ref{fig:four}(b). 
The data reported in Fig.~\ref{fig:six}(b) show indeed the same qualitative behavior reported in Fig.~\ref{fig:five}(b).
The only slight modification is that the drop of $R_A$ to zero, in the range $eV \gtrsim \Delta_0$, is less abrupt than in Fig.~\ref{fig:five}(b), due to a slight momentum mismatch  in the transmission channel  between GNR and S sides, arising when $U\neq 0$.

Before concluding we would like to comment on the double-spike structure of $R_{\rm A}$ at voltages $eV = \delta \pm U$, which is seen in both panels of Fig.~\ref{fig:six}. This is due to the opening of new channels for AR processes made available by the existence of the first-excited energy band (see Fig.~\ref{fig:four}).

\section{Conclusions}
\label{sect:conclusions}

In summary, we have studied Andreev reflection in graphene nanoribbon/superconductor hybrid junctions.

We have reported analytical expressions for the eigenfunctions of the tight-binding Hamiltonian describing 
the graphene nanoribbon, which are valid at arbitrary energy and carry explicitly a definite pseudoparity. 
The superselection rule stemming from this quantum number has two main implications on transport through the hydrid junction. 
For narrow nanoribbons with an even number of sites in the transverse direction we have found 
a complete suppression of Andreev reflection in a range of energies that is huge when the Fermi energy lies at the Dirac level. This implies zero conductance at subgap voltages, which can however be restored applying a finite gate potential, opening up potential technological applications of these hybrid junctions as electron transistors as well as nanorefrigerators. In contrast, in the case of narrow nanoribbons with an odd 
number of sites we have found perfect Andreev reflection at subgap voltages and an 
abrupt suppression of it at supergap voltages.

The study of non-ideal edges, realistic interfaces, and bulk disorder is beyond the scope of the present work and is postponed 
to a future publication~\cite{diego_dopo}.  The role of electron-electron interactions and/or next nearest-neighbor hopping, which also has 
not been addressed in the present work, can be qualitatively understood along the following lines. These effects have been shown to lead to the opening of a gap $\Delta_{\rm g}$ at the Dirac level~\cite{reviews,GNR_theory}, which is however typically much smaller than $\delta$. 
Our conclusions  thus remain valid for a large range of energies even when these effects are taken into account.

\acknowledgments 

We gratefully acknowledge Carlo Beenakker and Francisco Guinea for very useful discussions 
and suggestions. 
This work has been partly supported by the NANOFRIDGE EU Project and by the CNR-INFM ``Seed Projects''.


\begin{thebibliography}{77}
\bibitem{footnote_steplike}
	Notice that, since graphene honeycomb lattice is bipartite, a simplified model in which $\Delta_{\bm i}$ is assumed
	as a step-like function along the $\hat{\bm e}_{\bm x}$ direction would effectively imply a rapid variation of $\Delta_{\bm i}$
	also transversally along the two-site cells. Since such situation would not be realistic, we have adopted a smooth variation of $\Delta_{\bf i}$, over a length $\lambda \simeq 10 a$   along the $\hat{\bm e}_{\bm x}$ direction.
\bibitem{footnote_sqrt}
	The fact that $G$ vanishes in the range $\Delta_0 < eV < \sqrt{U^2 + \Delta^2_0}$ 
	can also be understood in terms of pseudoparity conservation.
\end{thebibliography}
\end{document}